\documentclass{article}
\usepackage{spconf,amsmath,graphicx,amssymb, multirow, booktabs, pifont, xcolor}
\ninept
\title{ Cross-Channel Attention-Based Target Speaker Voice Activity Detection: Experimental Results for M2MeT Challenge}

\name{Weiqing Wang$^1$, Xiaoyi Qin$^2$, Ming Li$^{1,2,}$\sthanks{Corresponding author: Ming Li}}
\address{$^1$Department of Electrical \& Computer Engineering, Duke University, Durham, NC 27708, USA\\
        $^2$Data Science Research Center, Duke Kunshan University, Kunshan 215316, PR China}
\email{ming.li369@duke.edu}

\begin{document}
%
\maketitle
\begin{abstract}
In this paper, we present the speaker diarization system for the Multi-channel Multi-party Meeting Transcription Challenge (M2MeT) from team DKU\_DukeECE. As the highly overlapped speech exists in the dataset, we employ an x-vector-based target-speaker voice activity detection (TS-VAD) to find the overlap between speakers. For the single-channel scenario, we separately train a model for each of the 8 channels and fuse the results. We also employ the cross-channel self-attention to further improve the performance, where the non-linear spatial correlations between different channels are learned and fused. Experimental results on the evaluation set show that the single-channel TS-VAD reduces the DER by over 75\% from 12.68\% to 3.14\%. The multi-channel TS-VAD further reduces the DER by 28\% and achieves a DER of 2.26\%. Our final submitted system achieves a DER of 2.98\% on the AliMeeting test set, which ranks 1st in the M2MET challenge.
\end{abstract}
\section{Introduction}
\label{sec:intro}

Meeting scenario is one of the most valuable scenarios for speech technologies and becomes increasingly important as the meeting is taking place every day. However, such a scenario is also one of the most challenging scenarios due to the highly overlapped speech and far-field signals with noise and reverberation. In addition, a fixed microphone array is usually equipped as the recording device in such a scenario, providing the multi-channel signals which do not only increase the difficulty but also improve the performance. 

As the conventional clustering-based speaker diarization system assumes that a segment only contains one speaker, it is not easy to recognize the overlapped speech without additional modules. Therefore, previous research focuses on reducing the error brought by overlapped speech, including the speech separation \cite{AHCSeg} as pre-processing, the target-speaker voice activity detection (TS-VAD) \cite{TSVAD} as post-processing, and the end-to-end neural diarization (EEND) as post-processing \cite{EENDPost} or directly producing the overlap-aware diarization results \cite{EENDEDA}.

Recently, the cross-channel attention has been successfully employed for multi-channel speech signals in speech enhancement \cite{CrossChannel1}, speech separation \cite{CrossChannel2, CrossChannel3}, speech recognition \cite{CrossChannel4} and speaker diarization \cite{CrossChannel5} and achieves good results on these tasks. Cross-channel attention can learn the non-linear contextual relationship across channels both within and across time steps. The outputs of the cross-channel attention are usually fused by global average pooling or concatenation for subsequent processing. 

In this paper, we present our speaker diarization system on the Alimeeting dataset for M2MeT
challenge \cite{M2MeT}. Considering that the average speech overlap ratio is over 30\% in this dataset, overlap detection is important to reduce the DER. Since TS-VAD achieved the state-of-the-art (SOTA) results in several diarization tasks \cite{TSVADDIHARD3, TSVADVOXSRC21}, we select it as our post-processing module for overlap detection and diarization results refinement. In addition, we employ cross-channel self-attention to fuse the speaker embedding of different channels during TS-VAD training to improve the performance further. 

The rest of this paper is organized as follows: Section 2 introduces the speaker embedding training. Section 3 presents the details of the diarization system. Section 4 gives the experimental results and discussion. Finally, Section 5 concludes this paper and presents the future work. 

\section{Deep Speaker Embedding Extraction}
\subsection{Dataset}
The experiments are conducted on the CN-Celeb dataset \cite{CNCeleb} and AliMeeting dataset \cite{M2MeT}. We adopt the two-stage training method to optimize the speaker embedding model. In the pre-training stage, the CN-Celeb dataset is employed as the training set. Next, the model is separately fine-tuned on two different datasets, one is the AliMeeting training set, and another is the combination of the CN-Celeb and AliMeeting training set. As the AliMeeting dataset does not provide the single-speaker utterances, we select all non-overlapped speech segments for speaker embedding training, where the segments shorter than 2 seconds are dropped. Finally, we build a trial set from the AliMeeting evaluation set to evaluate the performance of the speaker embedding system. The trial set contains 10692 trials from 25 speakers. 

\subsection{Data Augmentation}
We perform online data augmentation \cite{cai2020fly} with MUSAN dataset \cite{musan}. We only augment the speech with ambient noise for the background additive noise, as noise with speech is not allowed in this challenge. In addition, 40,000 simulated room impulse responses (RIRs) from small and medium rooms are used for the reverberation. To further enrich the training samples, we adopt the amplification and tempo (change audio playback speed but do not change its pitch) to audio signals. 

\subsection{Deep Speaker Embedding Model}
We employ the ResNet34 \cite{ResNet} as our speaker embedding model, and the structure of the model is the same as that in \cite{cai18_odyssey}. The encoding layer is based on global statistic pooling (GSP), and the dimension of the speaker embedding layer is 128. The ArcFace \cite{deng2019arcface} with a margin of 0.2 and softmax prescaling of 32 is used to train the speaker embedding model. During the evaluation phase, cosine similarity is employed for scoring.

\section{Multi-channel Speaker Diarization}
\subsection{Dataset}
We only use the Alimeeting dataset in the speaker diarization system. For the clustering-based system, the model is trained on the training set and evaluated on the evaluation set. For the TS-VAD model, we create a simulated dataset from the Alimeeting training set, and the simulation process is as follows:

\begin{enumerate}
    \item As each speaker in the AliMeeting dataset has a unique identification, we select all non-overlapped speech for each speaker from the AliMeeting training set for simulation. 
    \item Extract the labels from the transcript of the AliMeeting training set and remove all silence regions. 
    \item During the training stage, the simulated data is generated in an online manner, where we randomly choose a segment of the label and fill the active region with the continuous non-overlapped speech segments.
\end{enumerate}
The more detailed simulation process can be found in \cite{TSVADVOXSRC21}. Finally, the Alimeeting evaluation set is adopted as the validation and evaluation set. 

\subsection{Clustering-based System}
The clustering-based system is the same as that in \cite{LSTM}, where we use an LSTM-based network to extract the affinity matrix. First, we perform the uniform segmentation on all speech regions with a length of 1.28s and a shift of 0.64s. Next, the speaker embedding is extracted from all segments and augmented with Diac-augmentation \cite{Diac} with a probability of 0.8. The speaker embedding sequences are then fed to the network for training. The training details can be found in \cite{LSTM, TSVADVOXSRC21}. Finally, we employ spectral clustering to get the initialized diarization result, which is the input of the TS-VAD system. 

\begin{figure}[h]
    \centering
    \includegraphics[scale=1]{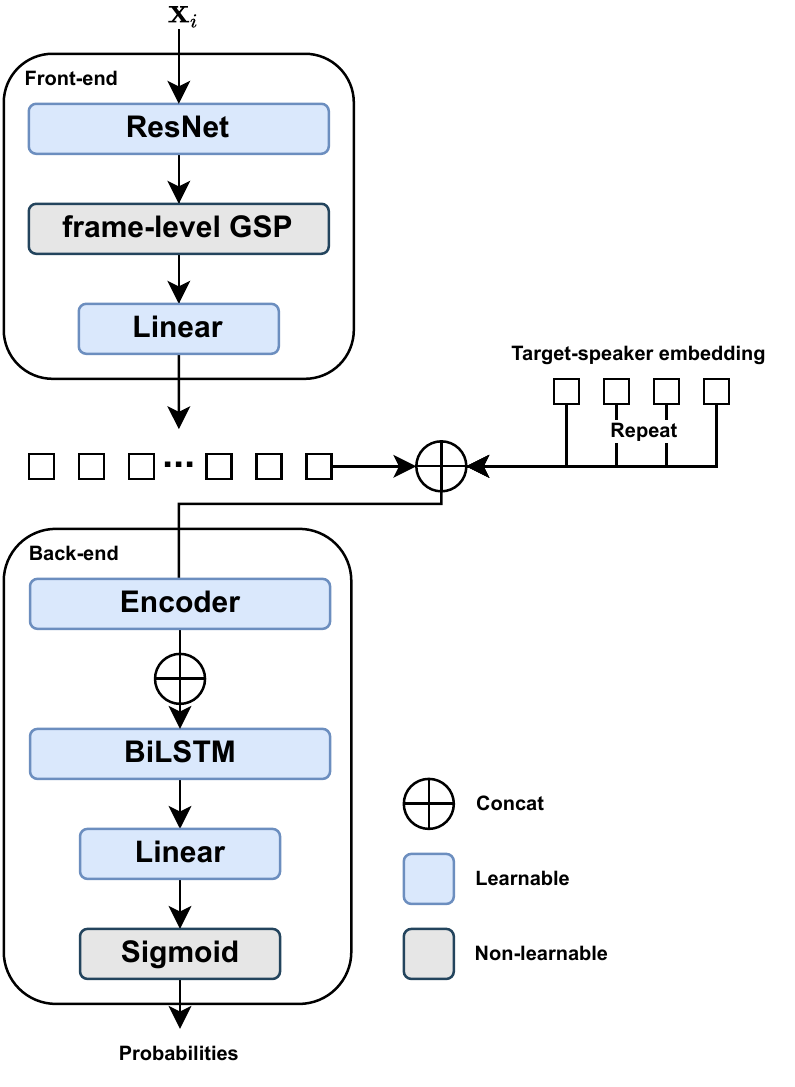}
    \caption{The architecture of the single-channel TS-VAD model}
    \label{fig:SC}
\end{figure}

\begin{figure}[h]
    \centering
    \includegraphics[scale=1]{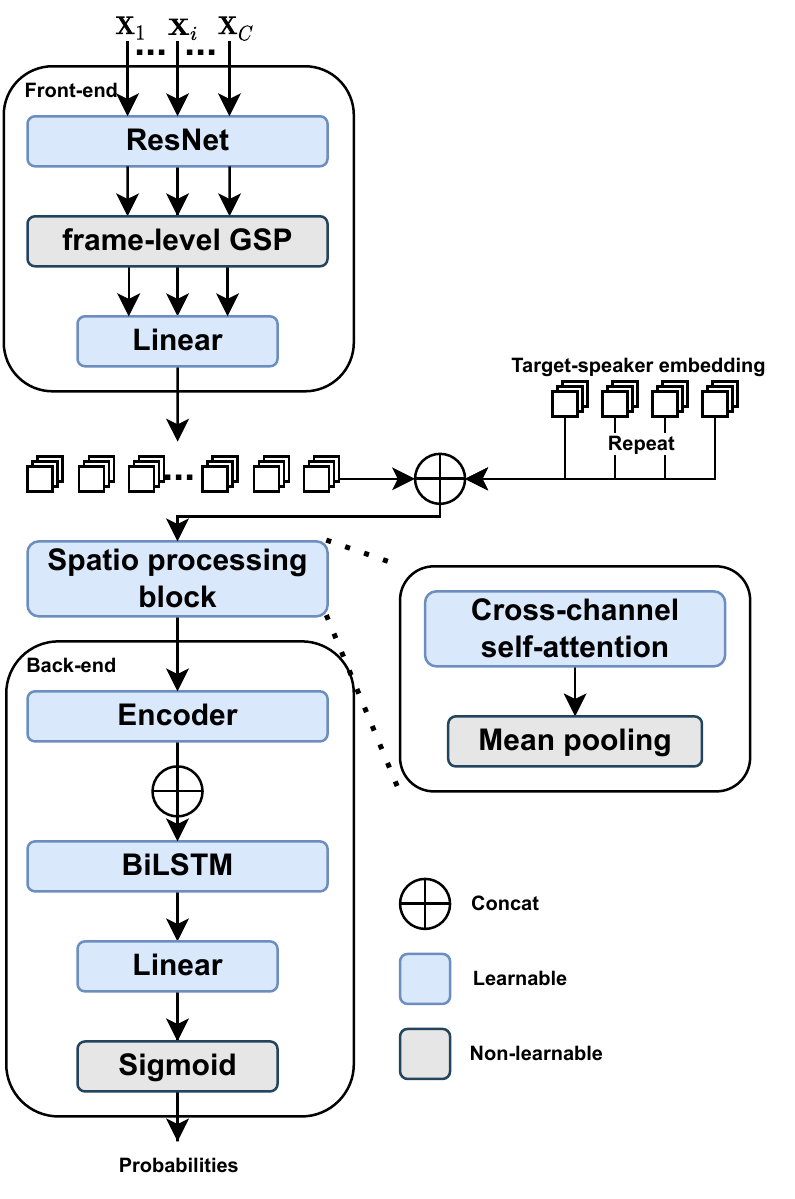}
    \caption{The architecture of the multi-channel TS-VAD model}
    \label{fig:MC}
\end{figure}
\begin{table*}[tp] 
  \caption{The performance of speaker embedding model.}
  \label{tab:spk_result}
  \centering
  \begin{tabular}[c]{llcccc}
    \toprule
     \multirow{2}*{\textbf{Training data}} & \multirow{2}*{\textbf{Training stage}} &  \multicolumn{2}{c}{\textbf{CN-Celeb trial}} &  \multicolumn{2}{c}{\textbf{AliMeeting trial}}  \\
     
     \cmidrule(lr){3-4} \cmidrule(lr){5-6}
   
      & & \textbf{EER[\%]}  & \textbf{mDCF$_{0.01}$} & \textbf{EER[\%]}  & \textbf{mDCF$_{0.01}$} \\
   \midrule
	 CN-Celeb  & Pre-train & 12.65 & 0.6751 & - & -     \\
     CN-Celeb + AliMeeting & Fine-tune & - & - & \textbf{3.199} & \textbf{0.353} \\
     AliMeeting & Fine-tune & - & - & 3.816 & 0.382 \\
     \bottomrule
     \end{tabular}
\end{table*}
\subsection{Single-channel TS-VAD}
Unlike the original TS-VAD \cite{TSVAD} that takes i-vector as target-speaker embedding, we use the deep speaker embedding extracted by ResNet to detect the target-speaker. Figure \ref{fig:SC} shows the architecture of our single-channel TS-VAD model. First, a ResNet, which has the same architecture as the speaker embedding model, extract the frame-level speaker embeddings. Next, the target-speaker embeddings are concatenated with the frame-level speaker embeddings. An encoder layer separately extract the detection state of each target-speakers. These detection states are concatenate and process by a BiLSTM to find the relationship between each speaker. Finally, a linear layer with a sigmoid function produce the final decision for each target speaker. The encoder can be a multi-layer BiLSTM or a Transformer encoder that can encode the contextual information across time, and the Transformer encoder is adopt in out experiments. 

The training step is as follows:
\begin{enumerate}
    \item Copy the parameters of the pre-trained speaker embedding model to the front-end model in the TS-VAD model. We keep the front-end model frozen and only train the back-end model on the simulated data for 10 epochs with a learning rate of $10^{-4}$. 
    \item Next, after the back-end model converges, we unfreeze the front-end model and jointly train the whole model for another 10 epochs with a learning rate of $10^{-4}$.
    \item Finally, we fine-tune the model on the AliMeeting training set for 200 epochs with a learning rate of $10^{-5}$. The 5 best models with the lowest validation loss are averaged as the final model for evaluation and inference. 
\end{enumerate}
The model is optimized by the Adam optimizer with binary cross-entropy loss. The input is 16s chunked waves, and the acoustic feature is 80-dim log Mel-filterbank energies (Fbank) with a frame length of 25ms and a frame shift of 10ms. The dimension of the output is a 4-dimensional vector that represents the existence probabilities for each speaker at each time step as the maximum number of speakers in the AliMeeting dataset is 4. We separately train a model for each channel, producing 8 single-channel TS-VAD models.

These three training steps are very important to obtain a good result in our experiments. We also try to directly train the whole model, but the model cannot converge for a very long time no matter we train from scratch or copy the parameters of the pre-trained speaker embedding model to the front-end ResNet. 

\subsection{Multi-channel TS-VAD}
The training set and training process are the same as that of the single-channel TS-VAD, and the only difference is the architecture and the input of the model. Figure \ref{fig:MC} shows the architecture of the multi-channel TS-VAD, where a cross-channel self-attention layer is employed on the concatenated embeddings to learn the cross-correlation between different channels. 

Given $C$-channels of Fbank sequence $\mathcal{X} = (\mathbf{X}_1, ..., \mathbf{X}_i, ...,\mathbf{X}_C)$ and the target-speaker embedding from $N$ target speakers $\tilde{\mathcal{S}} = (\tilde{\mathbf{S}}_{1}, ..., \tilde{\mathbf{S}}_{i}, ..., \tilde{\mathbf{S}}_{C})$, the corresponding target speaker decision is $\mathcal{Y} = (\mathbf{y}_1, ..., \mathbf{y}_t, ..., \mathbf{y}_T)$, where $\mathbf{X}_i \in \mathbb{R}^{T \times F}$ is of $T$ frames, $\tilde{\mathbf{S}}_{i}\in \mathbb{R}^{N \times D}$ is the D-dimensional speaker embedding from N target speakers, and $\mathbf{y}_t \in \{0, 1\}^N$ is the target speaker decision at time step t with dimension of N. Next, the front-end ResNet takes the $C$-channels of Fbank sequence as input and produce a frame-level speaker embedding sequence $\hat{\mathcal{S}} = (\hat{\mathbf{S}}_{1}, ..., \hat{\mathbf{S}}_{i}, ..., \hat{\mathbf{S}}_{C})$, where $\hat{\mathbf{S}}_{i} \in \mathbb{R}^{T \times D}$ is the frame-level speaker embedding  from the $\text{i}^{\text{th}}$ channel. Later, the target-speaker embedding is repeated T times and the frame-level speaker embedding is repeated N times. These two speaker embedings are concatenated at the embedding dimension: ${\mathcal{S}} = ({\mathbf{S}}_{1}, ..., {\mathbf{S}}_{i}, ..., {\mathbf{S}}_{C})$, where ${\mathbf{S}}_{i}\in \mathbb{R}^{T \times N \times 2D}$. 

The cross-channel self-attention takes the concatenated speaker embedding $\mathbf{S}_\text{in}$ as input:
\begin{align}
    \mathbf{S}_\text{in} &= \text{concat}(\mathcal{S}) \in \mathbb{R}^{T \times N \times C \times 2D} \\
    \mathbf{Q}^i &= \mathbf{W}_Q^i\mathbf{S}_\text{in} + \mathbf{b}_Q^i \\
    \mathbf{K}^i &= \mathbf{W}_K^i\mathbf{S}_\text{in} + \mathbf{b}_K^i \\
    \mathbf{V}^i &= \mathbf{W}_V^i\mathbf{S}_\text{in} + \mathbf{b}_V^i,
\end{align}
where $\mathbf{Q}^i$, $\mathbf{K}^i$ and $\mathbf{V}^i$ denotes the query, key and value matrices for the $\text{i}^{\text{th}}$ head. $\mathbf{W}^i \in \mathbb{R}^{E \times D}$ and $\mathbf{b}^i \in \mathbb{R}^{E}$ are the weight and bias for the $\text{i}^{\text{th}}$ head. Next, the scaled dot-produce attention is applied to the query, key and value:

\begin{align}
    \text{Attention}(\mathbf{Q}^i, \mathbf{K}^i, \mathbf{V}^i) = \text{softmax}\frac{\mathbf{Q}^i(\mathbf{K}^i)^\intercal}{\sqrt{n}}\mathbf{V}^i,
\end{align}
where $n=E$. Next, a positional-wise feed-forward layer with a ReLU activation is applied to generate the output, where layer norm and residual connections are employed between each layer. The output is denoted by $\mathbf{S}_\text{out} \in \mathbb{R}^{T \times N \times C \times 2D}$. Finally, the output is averaged over channel dimension by a global average pooling layer:

\begin{align}
    \mathbf{S}^\prime = \frac{1}{C}\sum_{i=1}^{C}S_{\text{out},i},
\end{align}
where $\mathbf{S}^\prime \in \mathbb{R}^{T \times N \times 2D}$. Finally, the back-end model processes this fused speaker embedding $\mathbf{S}^\prime$ in the same way as the singel-channel TS-VAD processes the concatenated speaker embedding. In our experiments, we employ a 2-layer and 2-head Transformer Encoder as the cross-channel self-attention layer.  

\subsection{Fusion}
We employ the DOVER-Lap \cite{dover} to fuse the systems that we mentioned above. 

\section{Experimental Results and Discussion}

\subsection{Speaker Embedding}
The performance of the speaker embedding model is reported in Table \ref{tab:spk_result}. As the model fine-tuned with CNCeleb + AliMeeting training set shows the best performance, it is adopted as the speaker embedding model for the speaker diarization task. 

\begin{table*}[h] 
  \caption{The performance of speaker diarization model on the AliMeeting evaluation set. For TS-VAD model, we report the DER of the 3rd round, where the input is the diarization results from the previous round. \\ $^*$ The result is from the MC-TS-VAD model initialized by ground-truth label.}
  \label{tab:dia_result}
  \centering
  \begin{tabular}[c]{lccccccc}
    \toprule
     \multirow{2}*{\textbf{Model}} &  \multicolumn{5}{c}{\textbf{AliMeeting evaluation set}} &  \multicolumn{2}{c}{\textbf{AliMeeting test set}}  \\
     
     \cmidrule(lr){2-6} \cmidrule(lr){7-8}
   
      & \textbf{MISS[\%]} & \textbf{FA[\%]} & \textbf{SpkErr[\%]} & \textbf{DER[\%]}  & \textbf{JER[\%]} & \textbf{DER[\%]}  & \textbf{JER[\%]} \\
   \midrule
     Clustering \\
     \ \ \ \  channel 1 & 10.6 & 1.2 & 1.9 & 13.80 & 27.03 & - & - \\
	 \ \ \ \  channel 2 & 10.5 & 1.4 & 1.0 & 12.86 & 24.19 & - & - \\
	 \ \ \ \  channel 3 & 11.0 & 1.2 & 1.0 & 13.27 & 25.56 & - & - \\
	 \ \ \ \  channel 4 & 10.8 & 1.4 & 0.9 & 13.13 & 24.77 & - & - \\
	 \ \ \ \  channel 5 & 10.9 & 1.3 & 1.1 & 13.38 & 25.06 & - & - \\
	 \ \ \ \  channel 6 & 10.7 & 1.3 & 1.0 & 12.97 & 23.86 & - & - \\
	 \ \ \ \  channel 7 & 10.8 & 1.3 & 1.0 & 13.08 & 23.67 & - & - \\
	 \ \ \ \  channel 8 & 10.6 & 1.2 & 0.6 & \textbf{12.68} & \textbf{23.09} & - & - \\
    DOVER-Lap fusion    & 11.6 & 0.8 & 0.8 & 13.23 & 24.89 & - & - \\
   \midrule
	 SC-TS-VAD \\
	 \ \ \ \  channel 1 & 2.5 & 1.0 & 0.7 & 4.12 & 12.43 & - & - \\
	 \ \ \ \  channel 2 & 2.6 & 1.1 & 0.6 & 4.26 & 12.82 & - & - \\
	 \ \ \ \  channel 3 & 2.8 & 1.0 & 0.4 & 4.21 & 12.78 & - & - \\
	 \ \ \ \  channel 4 & 2.5 & 0.9 & 0.5 & 3.93 & 12.56 & - & - \\
	 \ \ \ \  channel 5 & 2.6 & 0.9 & 0.4 & 3.95 & 12.08 & - & - \\
	 \ \ \ \  channel 6 & 2.5 & 0.9 & 0.4 & 3.90 & 11.86 & - & - \\
	 \ \ \ \  channel 7 & 2.3 & 0.9 & 0.4 & 3.61 & 11.55 & - & - \\
	 \ \ \ \  channel 8 & 2.4 & 0.8 & 0.3 & 3.49 & 11.28 & - & - \\
    DOVER-Lap fusion    & 2.3 & 0.5 & 0.3 & \textbf{3.14} & \textbf{11.08} & - & - \\
    \midrule
     MC-TS-VAD & 1.1 & 1.1 & 0.1 & \textbf{2.26} & \textbf{8.27} & \textbf{2.98} & - \\
     MC-TS-VAD$^*$ & 1.1 & 1.1 & 0.1 & 2.32 & 8.34 & - & - \\
     \midrule
     Baseline \cite{M2MeT} & - & - & - & 15.24 & - & 15.6 & - \\
     \bottomrule
     \end{tabular}
\end{table*}
\subsection{Speaker Diarization}
During the inference stage of the TS-VAD model, we first select all non-overlapped speech regions for each speaker based on the initialized results from the clustering-based system and extract the target-speaker embeddings. Next, we remove all silence regions and break the audio signals into 16s chunked waves with a 4s shift. After obtaining the probabilities from the TS-VAD model, we apply the median filtering with a window size of 7 to smooth the probability sequences. This results can also be the initialization of the next round of TS-VAD inference, and we infer it for 3 rounds. 

Table \ref{tab:dia_result} shows the results of all systems on the AliMeeting evaluation set. The missed speaker time (MISS), false alarm speaker time (FA), speaker error time (SpkErr), DER, and JER are reported for the evaluation set. In addition, for the clustering-based system and single-channel TS-VAD (SC-TS-VAD) system, the performance of each channel is reported, and the results of all channels are fused by DOVER-Lap. Results show that the model trained on the data of the 8th channel achieves the lowest DER, and fusion degrades the performance. For the SC-TS-VAD model, the fused system shows the best performance, and the MISS error is significantly reduced compared with the clustering-based method. The multi-channel TS-VAD (MC-TS-VAD) can further reduce the MISS error and achieve the lowest DER of 2.26\%. We also try to fuse this system with other systems, but the performance always goes worse. Therefore, our submitted system is this single system (MC-TS-VAD) without any fusion. 

In our experiments, we also find that the initialized results are not very important for TS-VAD in this challenge. The reason is that each of the recordings lasts over 30 minutes, which can produce a very long single-speaker segment for each speaker. If the number of speakers is correctly estimated, the TS-VAD model can always achieve the same results after several rounds even the initialized results are different. We also report the Multi-channel TS-VAD results initialized by the ground-truth label. It is interesting to note that even using the ground-truth label as initialization, the results are still similar to the results initialized by clustering-based results. Therefore, we do not explore the multi-channel method for the clustering-based system. 

\section{Conclusion and Future Work}

In this paper, we present our submitted system for the M2MeT challenge. As the dataset contains highly overlapped speech, most of the error of our clustering-based system is from the MISS error. Next, we employ the single-channel TS-VAD model to refine the diarization results and reduce the MISS error. Results show that the DER is significantly reduced by 75\%. In addition, we also apply cross-channel attention to the TS-VAD model to further improve the performance by 28\%. 

In the future, we will train a speaker embedding model on the multi-channel data and evaluate the performance of the diarization system. Also, we will perform some ablation studies on the cross-channel self-attention layer, e.g., applying the cross-channel self-attention on the detection states instead of the speaker embedding or increasing the number of parameters of the cross-channel self-attention.

\bibliographystyle{IEEEbib}
{\small
\bibliography{ref.bib}}

\end{document}